# Realistic Characterization and Modeling of Vehicle-Mounted Antenna Radiation Patterns Based on Large-Scale Full-Vehicle Measurements

Guotian Ji, Tianhao Wang, Changhao Xi, Guokai Jiang, Jiaxu Feng, Zhiqiang Yuan, and Wei Fan

***Abstract*—Vehicle-mounted antenna radiation patterns can be significantly altered after installation due to interactions with the vehicle body. However, existing studies mainly focus on simulations or isolated antenna characterization, while large-scale experimental investigations remain limited. This paper presents a comprehensive experimental analysis of vehicle-mounted antenna radiation patterns based on full-vehicle spherical near-field measurements. A large-scale dataset covering ten commercial vehicle platforms, multiple antenna mounting locations, and various wireless functions, including 4G, 5G, V2X, and GNSS, is established under a unified measurement framework. Based on the measured results, representative case studies and statistical characterization are performed to investigate the impacts of vehicle installation on antenna radiation behavior. The results reveal substantial variations across mounting locations and wireless functions, highlighting the strong influence of vehicle integration on realistic antenna performance. Furthermore, a novel statistical model is proposed to generate realistic vehicle-mounted antenna patterns using a small set of physically interpretable parameters. The obtained dataset, characterization results, and modeling approach provide a foundation for realistic antenna evaluation and wireless system simulations in vehicular communication applications.**

## I. Introduction

The rapid development of intelligent transportation systems and connected vehicles has significantly increased the demand for reliable vehicular wireless communication, while also imposing increasingly stringent requirements on automotive over-the-air (OTA) and full-vehicle antenna characterization [1]–[3]. Technologies such as 5G, C-V2X, Wi-Fi, and GNSS are becoming fundamental enablers for autonomous driving, cooperative perception, traffic coordination, and safety-critical services. In these systems, the radiation performance of vehicle-mounted antennas plays a critical role in determining communication reliability, coverage continuity, and link robustness.

In conventional antenna analysis and design, radiation characteristics are typically evaluated under free-space or simplified testing conditions, where the antenna behavior is mainly determined by its intrinsic structure [4]. However, practical vehicular antennas are integrated into electrically large and geometrically complex platforms. The vehicle body, including the roof structure, windshield, chassis, and surrounding components, introduces strong scattering, diffraction, reflection, and shadowing effects, and previous studies have shown that vehicle geometry, mounting position, roof integration, and glass-mounted configurations can substantially reshape the installed radiation behavior [5]–[8]. Consequently, the actual radiation pattern after vehicle integration can differ significantly from the isolated antenna pattern, leading to directional imbalance, severe radiation nulls, and coverage degradation. Therefore, the vehicle body should not be regarded merely as a passive supporting structure, but rather as an integral part of the overall radiating system.

Existing studies on vehicular antennas mainly focus on antenna design optimization, electromagnetic simulations, and measurement platform development. To support realistic vehicular antenna characterization, several advanced full-vehicle measurement facilities have been developed in recent years, including large far-field and hybrid test ranges [9], multi-probe spherical and planar near-field (SNF) systems [4], e.g., VISTA [10] and A-MST [12]. These facilities significantly improve measurement efficiency and enable realistic installed antenna characterization under controlled environments. Meanwhile, UAV-assisted measurement approaches [15], [16] and scaled-vehicle techniques [17] have also been explored to reduce measurement cost and improve flexibility. In addition to measurement platforms, several studies have investigated installed vehicular antenna performance through electromagnetic simulations and combined simulation-measurement frameworks [13], [14], [18], [19]. These works demonstrate that vehicle structures can strongly affect antenna radiation behavior and that mounting position optimization is important for practical deployment. Furthermore, measurement-based virtual-drive-testing methods have been proposed to connect realistic installed antenna patterns with V2X and vehicular communication performance under practical propagation conditions [20], [21].In particular, simulation-based studies [13], [14], [18] have analyzed the influence of vehicle platforms on an-

Guotian Ji and Tianhao Wang are with the College of Instrumentation and Electrical Engineering, Jilin University, Changchun 130026, Jilin, China; Guotian Ji is also with China Automotive Technology and Research Center Co., LTD, Tianjin 300300, China (e-mail: jigt25@mails.jlu.edu.cn; tianhao@jlu.edu.cn).

Changhao Xi and Wei Fan are with National Mobile Communications Research Laboratory, School of Information Science and Engineering, Southeast University, Nanjing 210096, China (e-mail: {213231572,weifan}@seu.edu.cn).

Guokai Jiang and Jiaxu Feng are with China Automotive Technology and Research Center Co., LTD, Tianjin 300300, China (email: {jiangguokai,fengjiaxu}@catarc.ac.cn).

Zhiqiang Yuan is with the Department of Electrical Engineering, Chalmers University of Technology, 41296 Gothenburg, Sweden (email: yuanzhiq@chalmers.se). Corresponding Author: Zhiqiang Yuan.

TABLE I
STATE-OF-THE-ART ANTENNA MEASUREMENT SYSTEMS

| Ref. | Platform | Measurement Method | Work Content |
|---|---|---|---|
| [4] | SATIMO hybrid system | Planar near-field measurement with gain calibration | Develops a hybrid antenna measurement system for improving measurement efficiency and accuracy. |
| [9] | Hemi-spherical NF/FF facility | Far-field and hemi-spherical near-field measurements | Presents a large-scale antenna test range supporting both NF and FF characterization. |
| [10] | VISTA vehicle test system | Multi-probe spherical near-field scanning | Introduces a rapid full-vehicle antenna measurement system for 3D radiation characterization. |
| [11] | EPG antenna facility | Outdoor compact-range far-field measurement | Describes a large outdoor antenna test facility for electrically large platforms. |
| [12] | A-MST multi-probe SNF system | Electronic spherical near-field scanning | Develops a fast multi-probe spherical near-field antenna measurement system. |
| [13] | In-vehicle antenna platform | SNF measurement combined with full-wave simulation | Investigates installed antenna performance in realistic vehicular environments. |
| [14] | CEM-based vehicle evaluation system | 3D electromagnetic simulation using CST | Builds a virtual framework for vehicle antenna placement evaluation. |
| [15] | UAV-based NF system | UAV-assisted near-field measurement | Proposes a UAV-assisted antenna measurement approach for flexible field testing. |
| [16] | SDR-UAV measurement system | Drone-path scanning and SDR processing | Develops a low-cost UAV-based outdoor antenna measurement system. |
| [17] | Scaled vehicle platform | SNF measurement with NF-FF transformation | Uses scaled vehicle models for experimental antenna verification. |
| [18] | Simulation | Full-wave and ray-tracing simulation | Studies integrated vehicle roof antennas using combined electromagnetic and propagation analysis. |
| [19] | Simulation | HFSS SBR simulation | Analyzes the installed radiation behavior of vehicle glass-mounted antennas. |

tenna coverage and propagation characteristics, while recent measurement-assisted investigations [19] further verified the impact of realistic vehicle structures on installed antenna performance.In addition to conventional range design, recent automotive OTA studies have emphasized several practical measurement challenges, including near-field facility comparison, chamber-floor effects, measurement repeatability, and angular sampling strategies for spherical near-field characterization [2], [3], [8], [22], [23]. Meanwhile, compact parametric representations of antenna radiation patterns have also been investigated to reduce the dependence on exhaustive full-pattern data and support efficient system-level simulations [24].

Nevertheless, most existing studies remain limited to simulation-based evaluations, measurement platform validation, virtual-drive-testing frameworks, compact antenna-pattern modelling, or single-case investigations [13], [14], [18]–[21], [24].A systematic and measurement-driven understanding of vehicle-induced radiation pattern distortion across different vehicle platforms, antenna locations, operating frequencies, and wireless functions remains insufficient. In particular, large-scale full-vehicle empirical studies capable of revealing statistical characteristics are still scarce due to the high complexity and cost of vehicular OTA measurements. To address these limitations, this paper presents a large-scale measurement-driven study of vehicle-mounted antenna radiation patterns based on full-vehicle spherical near-field measurements. The corresponding measurement dataset, statistical analysis framework, and realistic pattern characterization are developed to systematically investigate vehicle-induced radiation pattern distortion. The main contributions of this paper are summarized as follows:

- A large-scale full-vehicle antenna radiation measurement campaign is conducted using a unified high-resolution SNF measurement platform. A comprehensive dataset covering ten commercial vehicle types, multiple antenna mounting locations, wireless functions, and operating frequency bands is established.
- An empirical analysis of vehicle-mounted antenna radiation patterns is performed based on the measured dataset. The impacts of vehicle structure, mounting position, operating frequency, and antenna function on radiation behavior are characterized through representative case studies and statistical analyses.
- A novel statistical model is proposed to generate realistic vehicle-mounted antenna patterns using a small set of physically interpretable parameters. Results demonstrate the model can capture and reproduce the dominant characteristics of measured vehicular-mounted antenna patterns.

The remainder of this paper is organized as follows. Sec-tion II describes the measurement platform and experimental methodology. Section III introduces the statistical evaluation metrics. Section IV presents comparative analysis and discussion. Section V concludes the paper.

## II. Full-Vehicle Measurement Campaign

### A. Measurement Platform Overview

The experimental campaign was conducted at the Gaoyou Antenna Measurement Facility, Shenzhen, using a high-precision SNF measurement system, as illustrated in Fig. 1. The facility is designed for full-vehicle over-the-air (OTA) characterization and provides a fully anechoic indoor environment with low external electromagnetic interference and suppressed chamber reflections. The chamber dimensions are approximately 19.9 m × 16 m × 12.55 m, with a sliding door size of 3.6 m×3.6 m and an upper usable frequency of 40 GHz, accommodating vehicles up to 6 m in length. A nominal test distance of 6 m is employed for pattern acquisition. The measurement system utilizes a high-precision single-probe mechanical scanning arm with a radius of 6 m, together with an automated positioning turntable of 7 m diameter. The turntable supports a maximum load of 10 t (vehicle lift platform 3.5 t) and provides an angular positioning accuracy of $0.05^{\circ}$ over an elevation rotation range of $0^{\circ}$–$105^{\circ}$, which is essential for preserving the integrity of the $0.5^{\circ}$ high-resolution angular sampling strategy adopted throughout this study.

To achieve the ultra-high angular resolution required for this study, the platform utilizes a high-precision single-probe mechanical scanning system. Unlike multi-probe arrays, this single-probe architecture eliminates mutual coupling between sensors and allows for a continuous, high-fidelity characterization across a broad frequency spectrum from 400 MHz to 8 GHz (covering 5G NR n41/n78/n79, V2X, and GNSS). The system integrates an automated positioning turntable with an exceptional angular accuracy of $0.05^{\circ}$. This mechanical precision is the prerequisite for implementing a high-resolution sampling grid used in this research, to capture the rapid spatial fluctuations (ripples) and sharp nulls in the radiation patterns induced by complex scattering and reflections from the metallic vehicle body [23].

### B. Measurement Campaign

The measurement campaign was designed to acquire high-resolution radiation characteristics of realistic vehicle-mounted antenna systems under a unified and repeatable framework. All measurements were performed on complete vehicles rather than isolated antennas, thereby directly capturing the electromagnetic interactions between the antenna and the vehicle body. As summarized in Table II, the dataset covers ten commercial vehicle platforms, including sedans, SUVs, and MPVs, together with multiple antenna mounting locations such as rooftop modules, mirrors, bumpers, windshields, and interior regions. Various wireless systems are included, covering 4G, 5G, V2X, and GNSS applications over a wide frequency range from sub-GHz bands to 5.9 GHz. Both single-port and multi-port antenna configurations are considered to reflect practical vehicular communication systems.

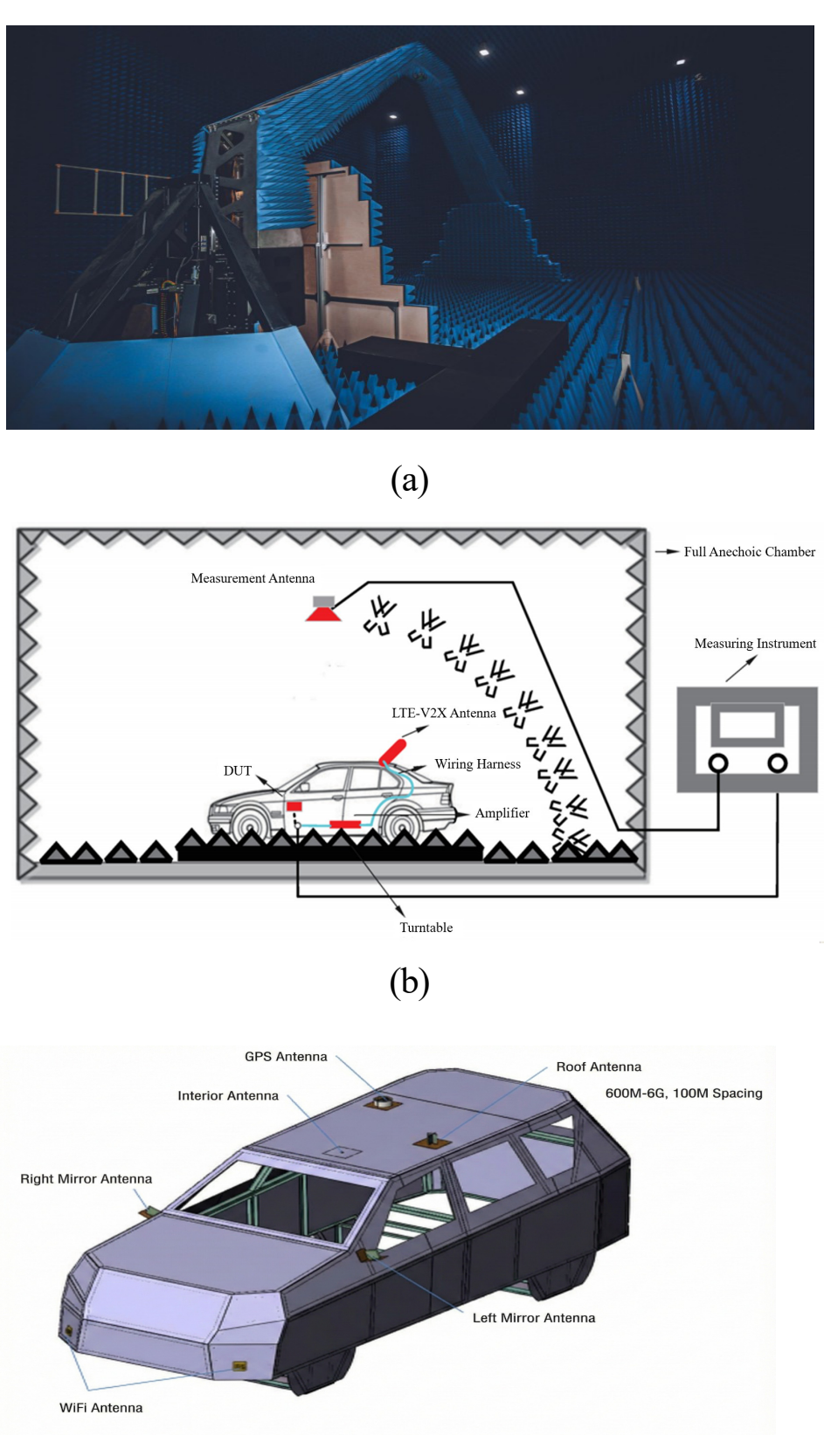

Fig. 1. Illustrations of the full-vehicle measurement campaign, including (a) photograph of the Gaoyou anechoic chamber, (b) measurement setup, and (c) antenna mounting locations on a vehicle.

To ensure cross-platform consistency, all vehicles were measured using the same chamber environment, hardware platform, angular sampling configuration, and post-processing procedure. A unified vehicle-centered coordinate system was adopted throughout the campaign, where the azimuth angle $\phi$ is referenced to the forward driving direction of the vehicle. Radiation patterns were acquired over the full azimuth range together with the upper hemispherical region using a dense angular sampling grid with resolutions of $0.5^{\circ}$ or $1^{\circ}$, depending on the operating frequency and measurement scenario. Prior to the measurement campaign, system calibration was performed using a standard gain horn antenna to verify the measurement accuracy and signal-chain consistency. The measured near-field data were subsequently transformed into far-field radiation patterns through near-field to far-field (NF-FF) transformation. The reconstructed far-field patterns were then used for the subsequent radiation analysis and statistical characterization of the vehicle-mounted antenna systems.

### C. Obtained Measurement Dataset

Based on the unified full-vehicle measurement campaign, a comprehensive dataset of vehicle-mounted antenna radiation patterns was obtained. The measurement scenarios are summarized in Table II, covering ten commercial vehicle platforms together with multiple antenna mounting locations, wireless functions, and operating frequency bands. The dataset includes rooftop, bumper, mirror, windshield, and interior-mounted antennas for 4G, 5G, V2X, and GNSS systems, thereby reflecting realistic vehicular antenna deployment conditions.

TABLE II
THE FULL-VEHICLE ANTENNA MEASUREMENT CAMPAIGN

| No. | Vehicle Type | Antenna Function | Mounting Position | Frequency Range | Step | Sampling Range |
|---|---|---|---|---|---|---|
| 1 | Mid-large SUV | 1.a C-V2X (2 ports) | Tailgate | 5900–5920 MHz | 0.5° | $\theta$: 0°–96°, $\phi$: 0°–360° |
| | | 1.b 5G (4 ports) | Rearview Mirror | 3300–5000 MHz | | |
| 2 | Compact SUV | 2. C-V2X | Sharkfin & Roof | 5850–5930 MHz | 0.5° | $\theta$: 0°–96°, $\phi$: 0°–360° |
| 3 | Mid-size Sedan | 3. C-V2X | Sharkfin & Roof | 5850–5930 MHz | 0.5° | $\theta$: 0°–96°, $\phi$: 0°–360° |
| 4 | Sedan | 4. C-V2X | Left & Right Rooftop | 5905–5925 MHz | 1° | $\theta$: 0°–90°, $\phi$: 0°–359° |
| 5 | Sedan | 5. C-V2X | Rearview Mirrors | 5905–5925 MHz | 1° | $\theta$: 0°–93°, $\phi$: 0°–359° |
| 6 | Mid-size Sedan | 6.a GNSS | Rear Rack | 1176.45–1605.63 MHz | 1° | $\theta$: 0°–96°, $\phi$: 0°–360° |
| | | 6.b 5G | | 897.5–4850.1 MHz | 0.5° | |
| 7 | Large-size MPV | 7.a 4G | Roof Sharkfin | 824–2170 MHz | 1° | $\theta$: 0°–96°, $\phi$: 0°–360° |
| | | 7.b 5G | | 3300–5000 MHz | 0.5° | |
| | | 7.c GNSS | | 1559.05–1576.42 MHz | 1° | |
| | | 7.d V2X | | 5855–5925 MHz | 0.5° | |
| 8 | Large-size SUV | 8.a 5G | Center Console | 3350.01–4942.01 MHz | 0.5° | $\theta$: 0°–100°, $\phi$: 0°–360° |
| | | 8.b GNSS | | 1176.45–1602 MHz | 1° | |
| 9 | Large-size Sedan | 9.a 5G | Roof Panel | 1710–2600 MHz | 1° | $\theta$: 0°–100°, $\phi$: 0°–360° |
| | | 9.b V2X | | 5905–5925 MHz | 0.5° | |
| 10 | Mid-size SUV | 10. C-V2X | Bumper | 5905–5925 MHz | 0.5° | $\theta$: 0°–100°, $\phi$: 0°–360° |

Representative examples of the measured radiation patterns are illustrated in Fig. 2 and Fig. 3. Specifically, Fig. 2 presents several representative three-dimensional radiation patterns measured under different vehicle platforms, antenna locations, and operating frequencies, demonstrating the significant diversity of installed radiation characteristics caused by vehicle-body interactions. Fig. 3 further shows the corresponding horizontal-plane and elevation-plane cuts, where strong directional imbalance, localized scattering lobes, and deep radiation nulls can be clearly observed in different installation scenarios.

The measured results reveal that the actual installed radiation patterns can differ substantially across vehicle structures, antenna locations, and communication functions, even within the same operating frequency band. These observations further indicate that realistic vehicle-mounted antenna behavior cannot be accurately characterized using isolated antenna measurements alone.

Overall, the obtained dataset provides a large-scale empirical foundation for investigating vehicle-induced radiation pattern distortion, statistical radiation characteristics, and realistic installed antenna behavior. The dataset further supports subsequent analysis, modeling, and system-level evaluation for practical vehicular communication systems.

## III. CHARACTERIZATION METRICS FOR VEHICLE-MOUNTED ANTENNA PATTERN ANALYSIS

A statistical characterization of vehicle-mounted antenna radiation patterns is expected using the large-scale measured dataset to demonstrate the realistic radiation performance. Compared with isolated antenna measurements, the installed radiation patterns exhibit strong spatial fluctuation and blockage-induced distortion. To address this issue, several basic statistical metrics are introduced here to evaluate the measured radiation patterns from the perspectives of omnidirectional radiation uniformity, directional radiation gain, and service region coverage. These metrics aim to quantitatively analyze vehicle-induced radiation pattern distortion across different vehicle platforms, antenna locations, operating frequencies, and wireless functions.

### *A. Horizontal Omnidirectionality Metrics*

Due to vehicle-body blockage and scattering, antennas originally designed with omni-directional radiation may no longer preserve this property after vehicle integration. Therefore, two basic metrics, namely horizontal gain standard deviation $\sigma_{\mathrm{H}}$ and horizontal pattern range (HPR), are introduced to evaluate the horizontal omni-directional radiation pattern.

The horizontal gain standard deviation $\sigma_{\mathrm{H}}$ quantifies the azimuthal gain fluctuation and is defined as

$$\sigma_{\mathrm{H}} = \sqrt{\frac{1}{N}\sum_{i=1}^{N}\left(G_{\theta,i} - \bar{G}\right)^2}, \tag{1}$$

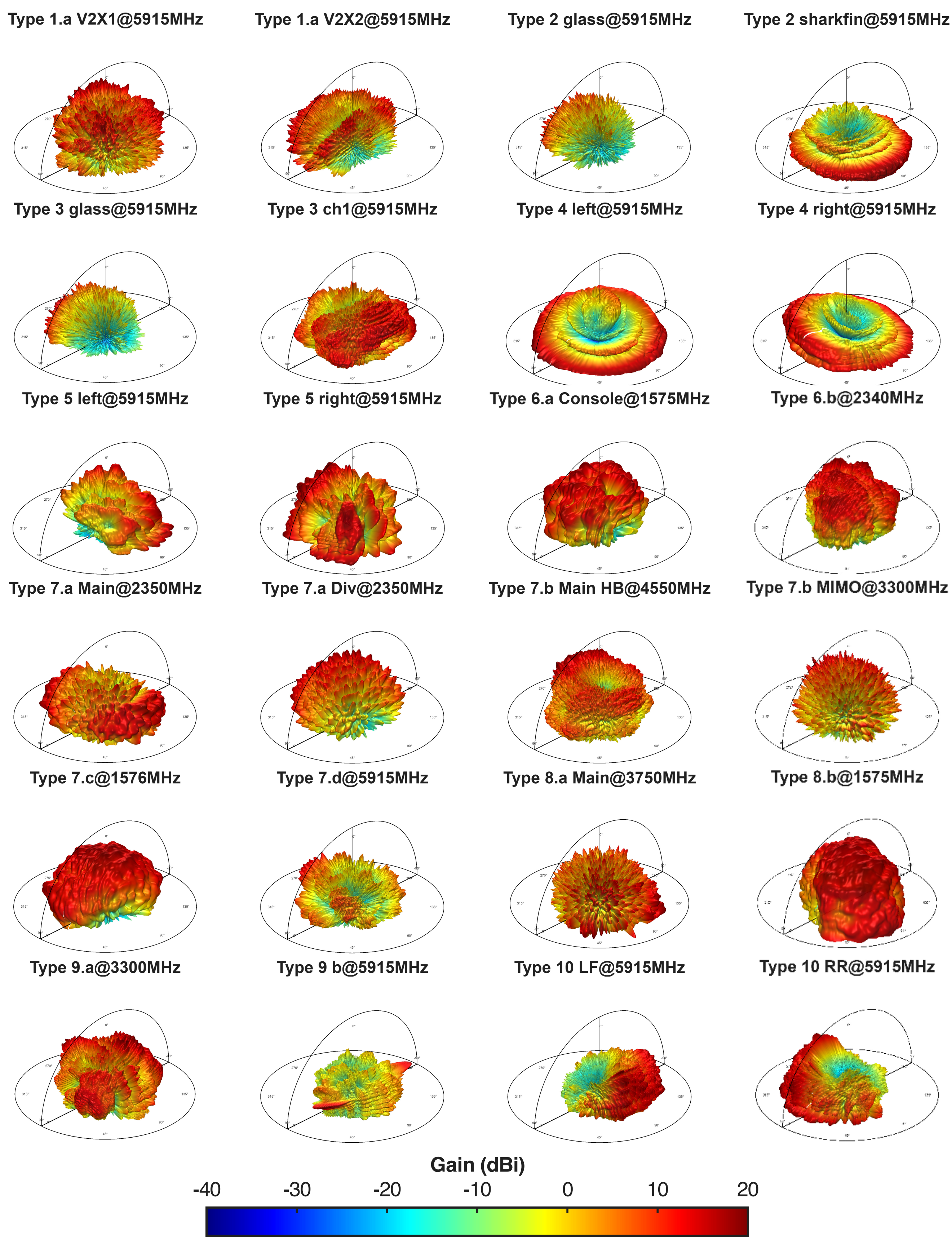

Fig. 2. Partial measured three-dimensional patterns of vehicle-mounted antennas for demonstration, from different vehicle types, locations, and frequencies.

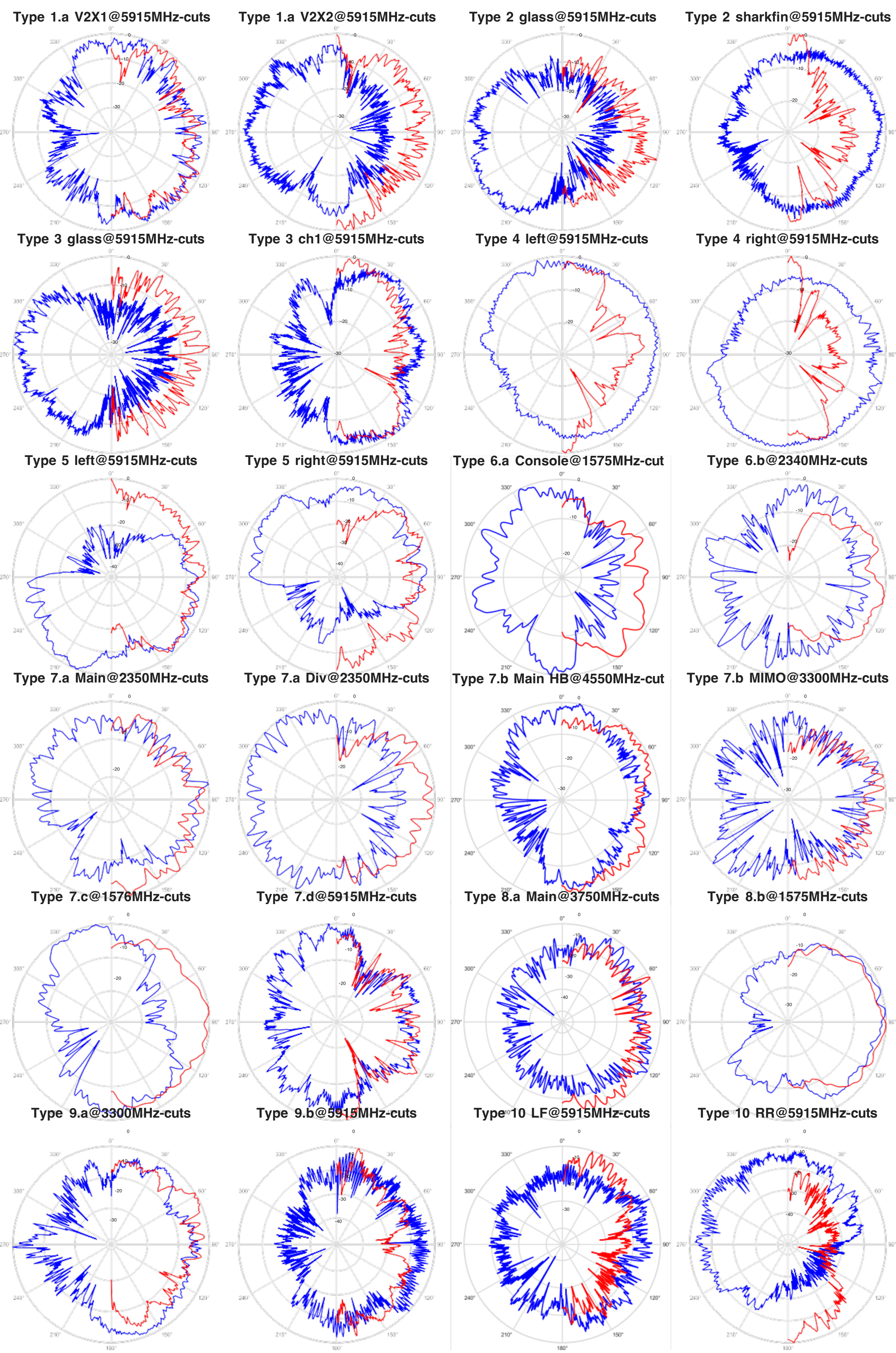

Fig. 3. H-plane (blue) and E-plane (red) cuts of measured antenna patterns.

where $G_{\theta,i}$ is the average gain at the $i$-th azimuth angle over the selected elevation range, $\bar{G}$ is the mean value of $G_{\theta,i}$ over all azimuth samples, and $N$ is the number of azimuth sampling points. A smaller $\sigma_{\mathrm{H}}$ indicates a more uniform horizontal radiation pattern.

While $\sigma_{\mathrm{H}}$ describes the overall azimuthal fluctuation, HPR characterizes the horizontal dynamic range and highlights severe radiation variations such as deep nulls caused by vehicle-body shadowing. It is defined as the difference between the 95th and 5th percentile gains of the azimuthal distribution:

$$\mathrm{HPR} = P_{95} - P_5. \tag{2}$$

A smaller HPR indicates a smoother and more omnidirectional horizontal radiation pattern.

### B. Partial Average Gain

Recommended by 3GPP and CTIA, Partial Average Gain (PAG) evaluates the aggregated radiation capability over a specified service angular region, providing an evaluation metric for the specific directivity of the antenna. For an angular region defined by $\theta_A \leq \theta \leq \theta_B$ and $\phi_A \leq \phi \leq \phi_B$ (divided into $p$ and $q$ intervals, respectively), PAG is computed as a weighted spatial average:

$$\mathrm{PAG} = \frac{\left(\frac{\mathrm{cut}_0 + \mathrm{cut}_p}{2} + \sum_{i=1}^{p-1} \mathrm{cut}_i\right)}{q\left(\frac{\sin\theta_0 + \sin\theta_p}{2} + \sum_{i=1}^{p-1} \sin\theta_i\right)} \tag{3}$$

where the azimuthally summed component $\mathrm{cut}_i$ is:

$$\mathrm{cut}_i = \sum_{j=0}^{q} \left[G_\theta(\theta_i, \phi_j) + G_\phi(\theta_i, \phi_j)\right] \sin\theta_i \tag{4}$$

Here, $G_\theta$ and $G_\phi$ are the orthogonal polarization gains, and the weighting factor $\sin\theta_i$ accounts for the spherical coordinate surface element.

### C. Upper-Hemisphere Efficiency (UHE)

Upper-Hemisphere Efficiency (UHE) quantifies the normalized radiation capability within the upper hemisphere, and is particularly useful for evaluating antenna functions that rely on radiation toward elevated directions, such as GNSS reception or links to cellular base stations. It is defined as

$$\mathrm{UHE} = \frac{1}{4\pi}\int_0^{2\pi}\int_0^{\pi/2} G_{\mathrm{lin}}(\theta,\phi)\sin\theta\, d\theta\, d\phi \times 100\%. \tag{5}$$

where $G_{\mathrm{lin}}(\theta,\phi)$ denotes the total realized gain in linear scale, and the factor $1/(4\pi)$ normalizes the integrated radiation over the full solid angle of an isotropic radiator.

## IV. Results and Modeling Analysis

This section presents the measured radiation-pattern results and their statistical characterization. Representative case studies are first analyzed to illustrate key vehicle-induced effects on antenna radiation patterns, including mounting-location dependence, antenna integration effects, and function-specific coverage characteristics. The complete dataset is then statistically characterized using $\sigma_H$, HPR, PAG, and UHE. Finally, a measurement-driven model is proposed for generating realistic vehicle-mounted antenna patterns.

### A. Specific Case Studies

*1) The Effect of Installation Locations:* Fig. 4 compares the radiation patterns of antennas with the same function but different mounting positions to illustrate the impact of installation locations of antennas on the vehicle. For Vehicle Type 2, Fig. 4(a) and Fig. 4(b) show the measured patterns of the glass-integrated antenna and the roof-mounted sharkfin antenna, respectively, whose installation locations are illustrated in 1(c). Due to the inclined glass surface and the nearby vehicle-body blockage, the glass-integrated antenna exhibits a more irregular pattern with stronger local fluctuations, and its main radiation is biased toward the forward side of the glass while the rear side is partially shadowed. In contrast, the sharkfin antenna presents a smoother and more omnidirectional horizontal pattern, which can be attributed to the more open radiation environment above the roof plane. This observation is also reflected by the statistical metrics: the sharkfin antenna has slightly lower $\sigma_H$ and HPR values, indicating better horizontal omnidirectionality, while its PAG is improved by about 0.8 dB, suggesting a higher average gain within the selected service region.

A similar comparison is made for the bumper-mounted antennas of Vehicle Type 10, as shown in Fig. 4(c) and Fig. 4(d). The left-front bumper antenna maintains a relatively open radiation region, whereas the right-rear bumper antenna suffers from more pronounced low-gain regions due to stronger cabin shadowing, rear-body blockage, and local scattering. Quantitatively, $\sigma_H$ and HPR increase from 6.12 dB and 20.22 dB to 9.20 dB and 28.67 dB, respectively, indicating more severe horizontal pattern distortion. Meanwhile, the PAG decreases from 9.72 dB to 6.07 dB, showing that the stronger blockage not only degrades the horizontal omnidirectionality but also reduces the average radiation capability in the selected service region.

*2) Integrated Pattern Analysis:* Since vehicle-body blockage can degrade the omnidirectional coverage of a single antenna, a common practical approach is to deploy multiple antennas around the vehicle and combine their radiation patterns. This case investigates the selection-combined V2X radiation patterns of distributed antennas and analyzes their horizontal omnidirectionality.

Fig. 5 shows the combined radiation patterns of two antennas mounted at different locations for Vehicle Types 4 and 10 at 5915 MHz. The combined pattern is obtained by selecting the maximum gain of the two antennas at each angular sample. As shown in Fig. 5(a) and 5(c), compared with the individual patterns, the combined results exhibit fewer deep nulls and more uniform azimuthal coverage, indicating improved omnidirectionality. This effect is particularly evident for Vehicle Type 10 as shown in Fig. 5(b) and 5(d), where blockage affecting one bumper-mounted antenna is compensated by the other mounted in another direction. The results further indicate the impact of the vehicle body on antenna patterns, and demonstrate that distributed antenna deployment can effectively mitigate vehicle-induced blockage and improve coverage continuity.

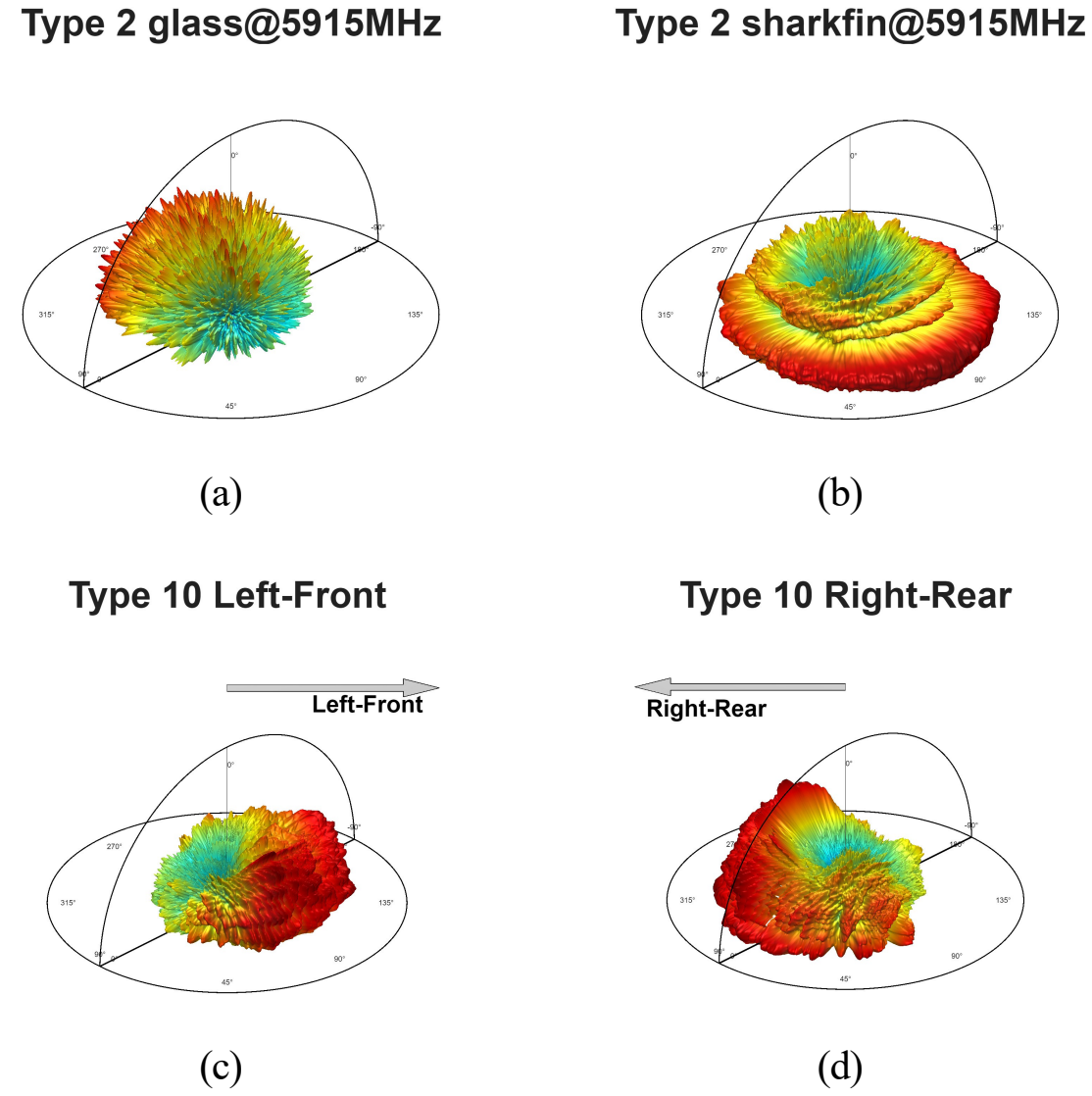

Fig. 4. Representative radiation patterns illustrating vehicle-induced distortion: antennas mounted on (a) the windshield and (b) the roof of Vehicle Type 2, and on (c) the left-front bumper and (d) the right-rear bumper of Vehicle Type 10.

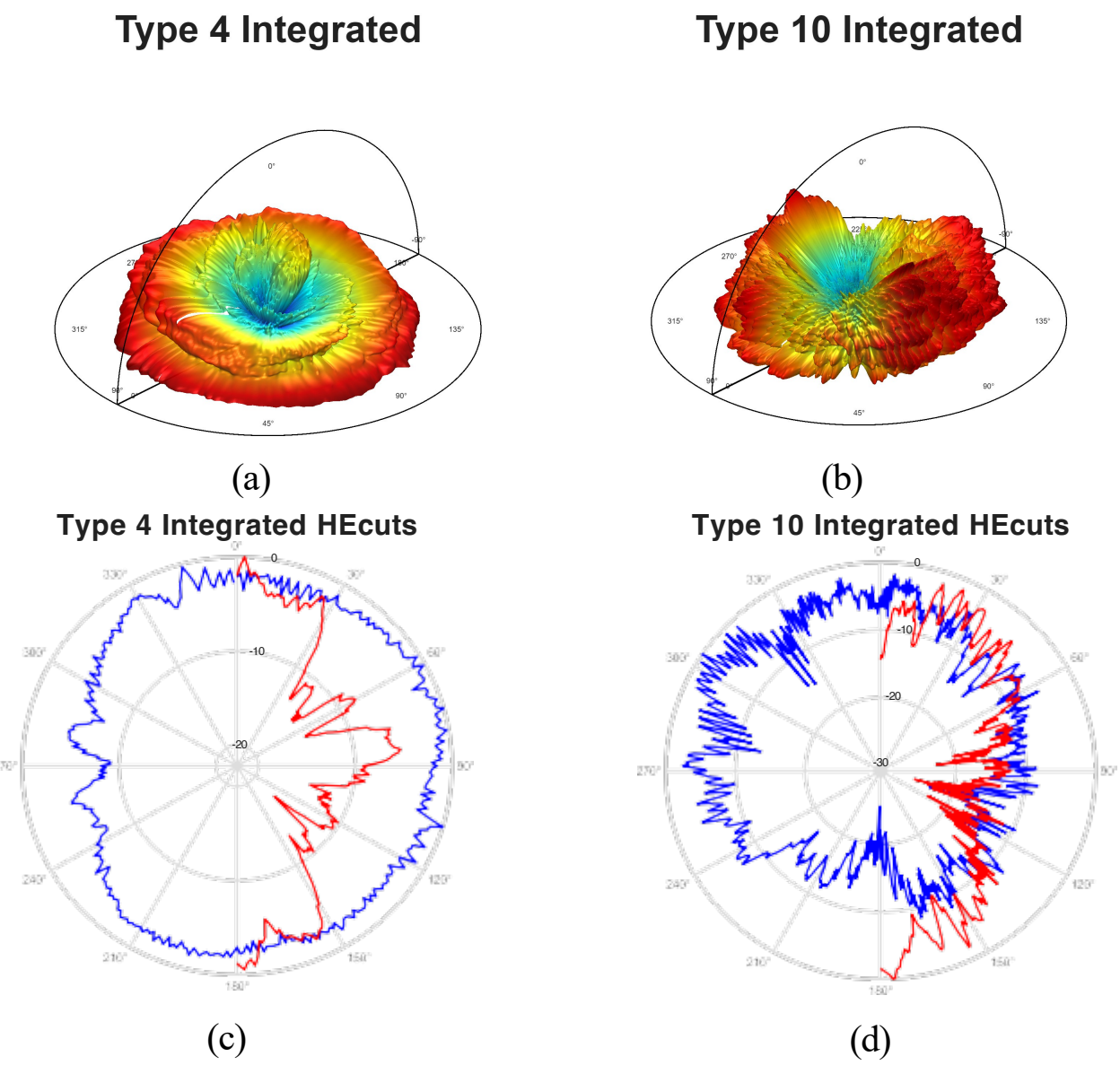

Fig. 5. Combined radiation patterns of two antennas mounted at different locations on Vehicle Types 4 (a,c) and 10 (b,d), obtained from full-vehicle measurements.

*3) Function-Specific Radiation Analysis:* Vehicle-mounted antennas serve different wireless functions and therefore exhibit distinct radiation requirements. GNSS antennas primarily require upper-hemisphere coverage for satellite reception, whereas V2X and 5G antennas rely on near-horizon radiation to support vehicular and terrestrial communication links. Fig. 6 compares representative radiation patterns for GNSS, V2X, and 5G antennas. These differences are clearly reflected in the measured patterns. The GNSS antenna concentrates radiation toward the upper hemisphere, while the V2X and 5G antennas emphasize horizontal coverage and are more sensitive to azimuthal nulls and coverage discontinuities. Quantitatively, the GNSS antenna achieves a UHE of 30.23%, indicating strong upper-hemisphere coverage, whereas the V2X and 5G antennas exhibit HPR values of 15.22 dB and 19.23 dB, respectively, reflecting their horizontal radiation variation. These results demonstrate that no single metric can adequately characterize all antenna functions. Instead, horizontal metrics such as $\sigma_H$ and HPR should be jointly considered with service-region metrics such as PAG and UHE to provide a function-aware evaluation of vehicle-mounted antenna performance.

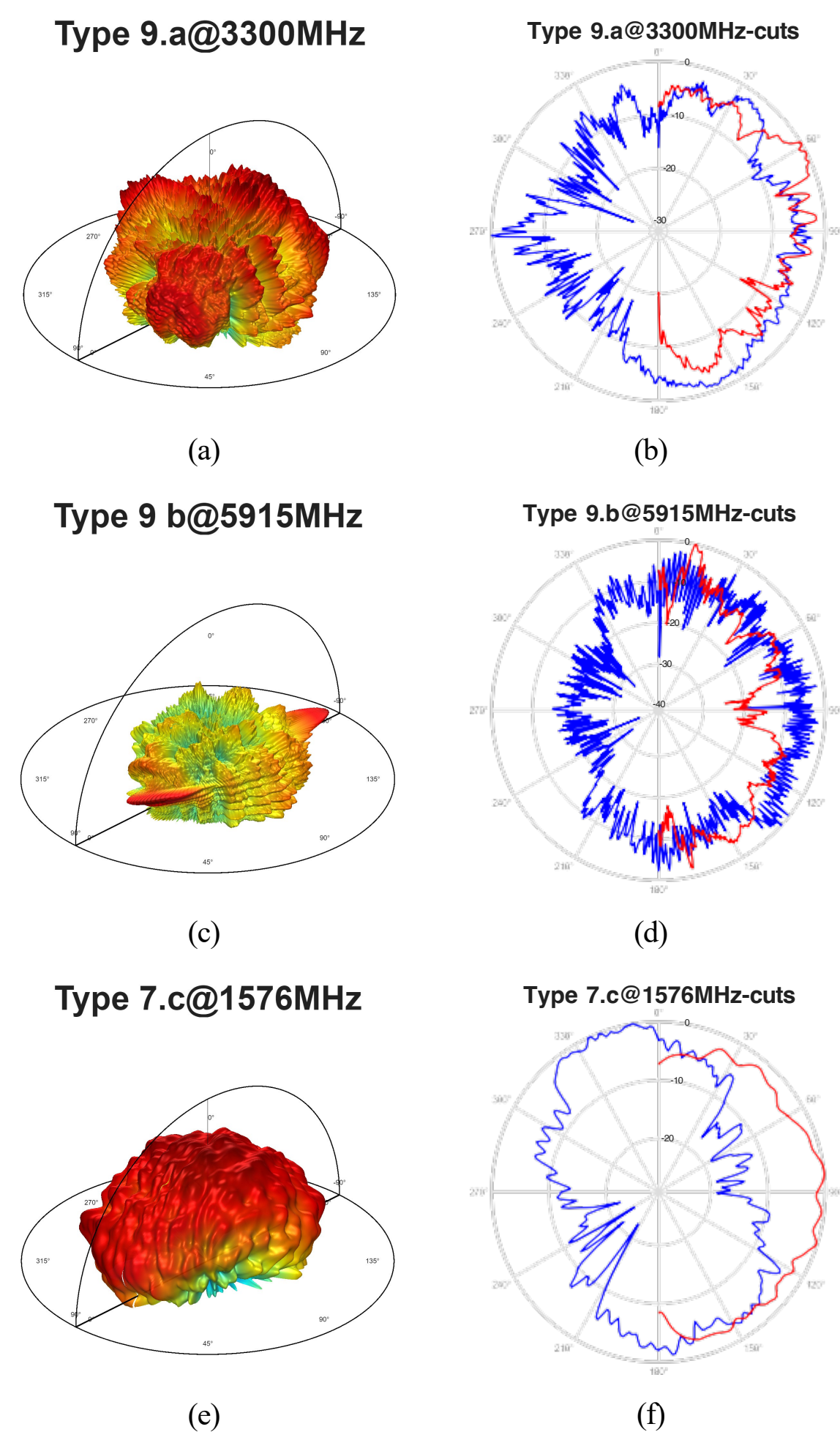

Fig. 6. Measured antenna patterns for different functions, including (a,b) 5G, (c,d) V2X, and (e,f) GNSS.

## B. Statistical Characterization

This subsection statistically characterizes the measured vehicle-mounted antenna radiation patterns using the complete dataset. The antennas are grouped according to their wireless functions, namely 4G/5G, GNSS, and V2X/C-V2X, while measurements at different frequency points are treated as independent samples within each category.

Fig. 7 shows the cumulative distribution functions (CDFs) of the horizontal gain deviation $\sigma_H$ and the HPR. The mean and variance of $\sigma_H$ are calculated as 4.98 and 0.31 for 4G/5G antennas, 4.53 and 0.86 for GNSS antennas, and 4.33 and 0.72 for V2X/C-V2X antennas, respectively. The corresponding means and variances of HPR are 10.50 and 11.18, 14.01 and 7.86, and 13.93 and 9.98, respectively. The largely overlapping $\sigma_H$ distributions suggest that the average level of horizontal gain fluctuation is broadly similar across different vehicular communication functions. By contrast, the wider spread of the HPR distributions for GNSS and V2X/C-V2X indicates greater variability in azimuthal coverage. This result highlights the strong influence of vehicle integration on installed radiation behavior, where vehicle-body blockage, scattering, and mounting-location effects can introduce significant directional nulls and coverage imbalance in certain deployment scenarios.

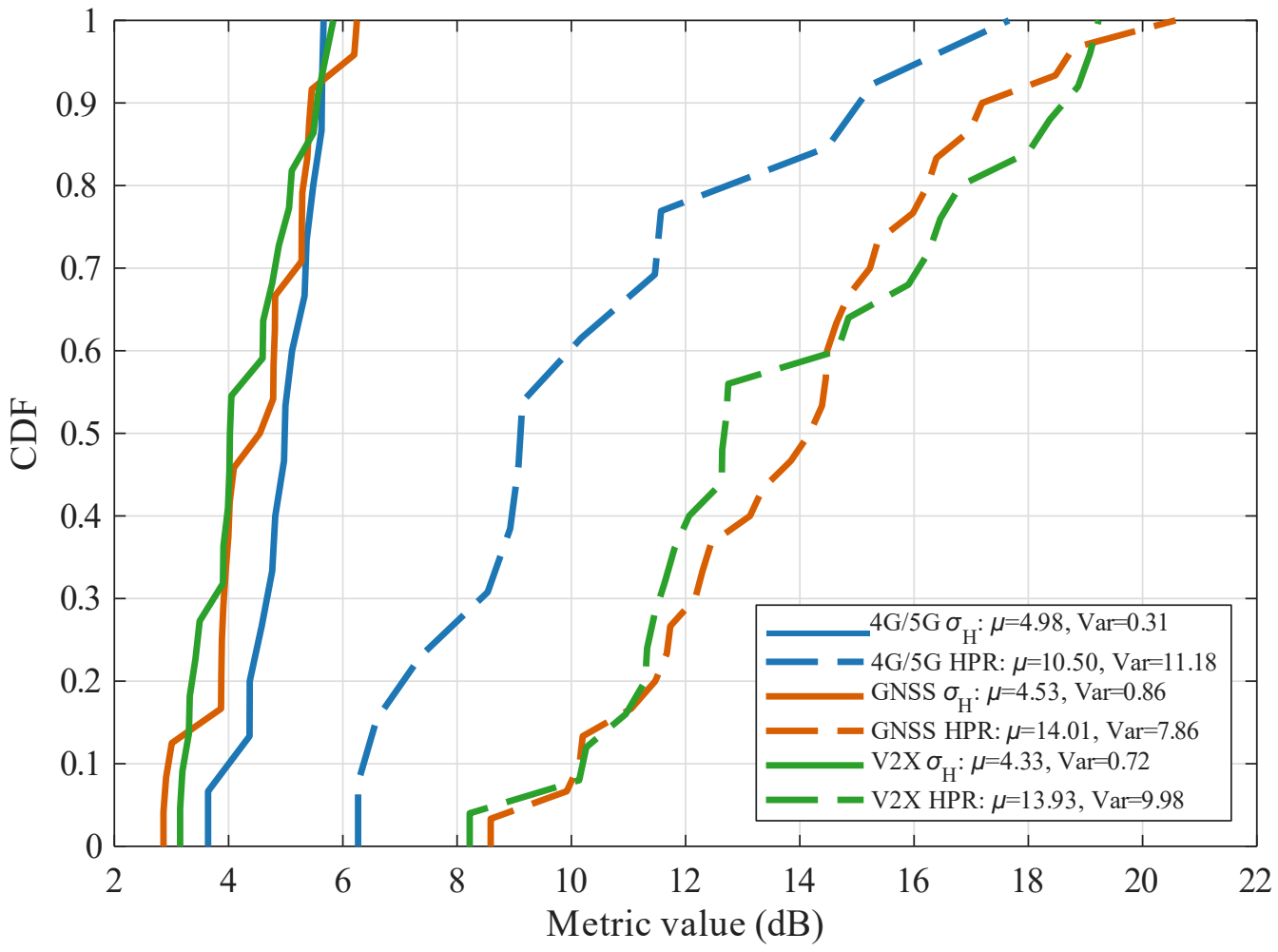

Fig. 7. CDFs of the horizontal gain deviation and HPR for the measured 4G/5G, GNSS, and V2X/C-V2X antennas.

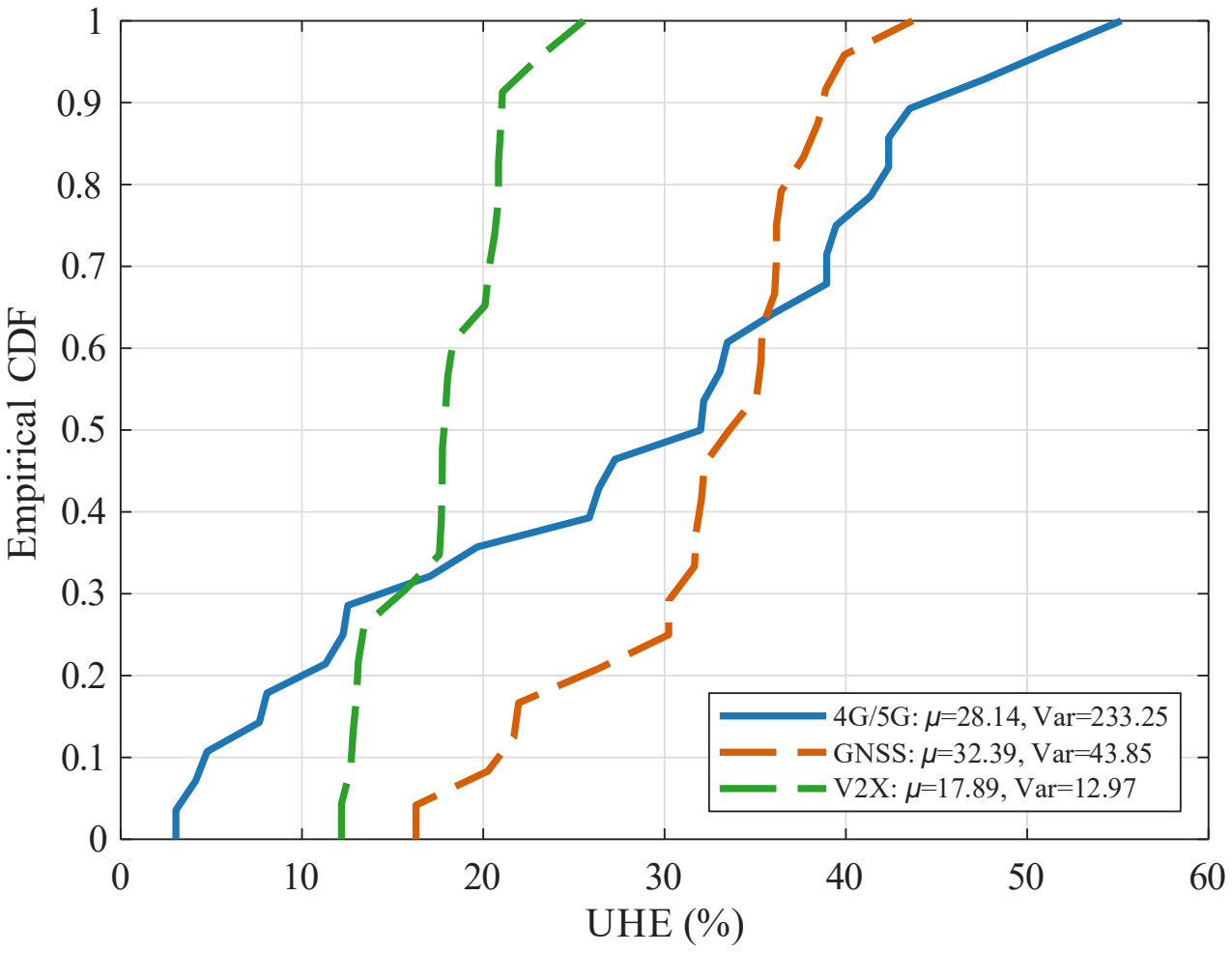

Fig. 9. CDFs of UHE for the 4G/5G, GNSS, and V2X/C-V2X antennas.

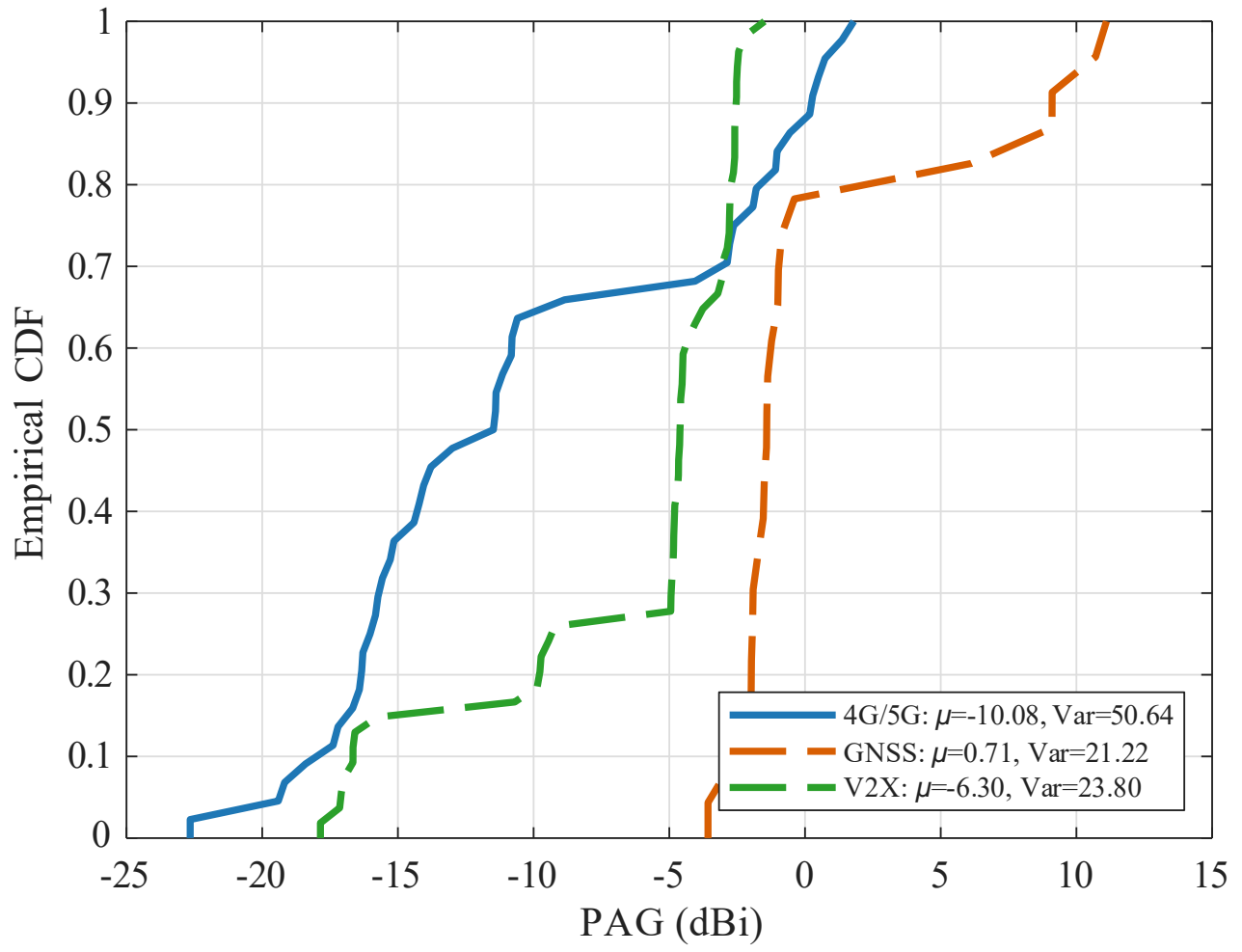

Fig. 8. CDFs of PAG for the 4G/5G, GNSS, and V2X/C-V2X antennas.

Fig. 8 shows the PAG distributions for the three function groups. The GNSS antennas exhibit the highest mean PAG of 0.71 dBi, indicating stronger average radiation within their service region. In contrast, the 4G/5G and V2X/C-V2X antennas have mean PAG values of -10.08 dBi and -6.30 dBi, respectively. The relatively broad distributions observed across all functions suggest that PAG is strongly influenced by both antenna installation and vehicle-body interactions, resulting in substantial variation in service-region coverage performance.

Fig. 9 presents the UHE distributions. As expected, the GNSS antennas achieve the highest mean UHE of 32.39%, reflecting their emphasis on upper-hemisphere coverage for satellite reception. The 4G/5G antennas exhibit a comparable mean UHE of 28.14% but a significantly larger spread, indicating stronger dependence on installation conditions. In comparison, the V2X/C-V2X antennas show a lower mean UHE of 17.89%, consistent with their focus on near-horizon communications. These results demonstrate that vehicle integration affects not only horizontal coverage characteristics but also elevation-domain radiation efficiency.

## C. *Vehicle-Mounted Antenna Modelling*

The preceding results demonstrate that the radiation behavior of a vehicle-mounted antenna is determined not only by the antenna itself, but also by its interaction with the vehicle after installation. While the statistical metrics introduced above provide a quantitative characterization of these effects, efficient pattern models are still needed for system-level vehicular simulations [25]–[27]. Therefore, based on the measured dataset, this subsection proposes a statistical model for realistic horizontal radiation patterns that captures the dominant effects of vehicle integration. Note that the proposed model is not intended to reproduce individual measurements. Instead, it statistically generates realistic vehicle-mounted antenna patterns using physically interpretable parameters.

Based on the measured dataset, vehicle-mounted antenna patterns are characterized by three key features: the baseline gain level, the severity of vehicle-induced distortion, and the dominant blockage direction. Correspondingly, three physically interpretable parameters, namely $G_0$, $Q$, and $\phi_B$, are extracted from the measured patterns, as summarized in Table III. Based on an empirical analysis of the measured dataset with these parameters and an understanding of the classical antenna theory [28], the horizontal radiation pattern of vehicle-mounted antennas can be modeled as,

$$\hat{G}(\phi) = G_0 + \alpha Q \cos(\phi - \phi_B - \pi) - (1-\alpha) Q \exp\left(-\frac{d^2(\phi, \phi_B)}{2\sigma_B^2}\right) + R(\phi), \quad (6)$$

where $R(\phi)$ denotes a stochastic residual component that accounts for small-scale angular ripples caused by local scattering and other irregular vehicle-body interactions, modeled as a zero-mean Gaussian random Fourier series. The cosine term represents large-scale directional bias, and the Gaussian notch models the main shadowing region caused by the vehicle body. The coefficient $\alpha$ balances directional bias and blockage-induced suppression, while $\sigma_B$ controls the angular width of the notch. Note that as summarized in Table III, the adopted parameters provide a physical link to antenna installations: $G_0$ represents the baseline radiation level, $Q$ quantifies the strength of vehicle-induced distortion, and $\phi_B$ specifies the dominant blockage direction.

TABLE III
PARAMETERS FOR STATISTICAL ANTENNA PATTERN GENERATION

| Parameter | Extraction Method | Role in Model |
|---|---|---|
| $G_0$ | Median of smoothed horizontal gain | Baseline gain level. |
| $Q$ | HPR, i.e., $P_{95} - P_5$ | Vehicle-induced distortion strength. |
| $\phi_B$ | Direction of deepest notch | Dominant blockage direction. |

For demonstration, we implement the proposed model for a selected V2X antenna, i.e., the Vehicle Type 10 left-front bumper antenna at 5915 MHz. The parameters calculated from the measured pattern are $G_0$ = 3.0 dB, $Q$ = 12.2 dB, and $\phi_B$ = 229.5°. The modeling results are shown in Fig. 10, where the measured pattern is compared with several model realizations (different realizations of $R(\phi)$). As shown, the generated patterns reproduce the observed angular fluctuations and show good agreement with the measured radiation trend. This indicates the capability of the proposed model to generate realistic vehicle-mounted antenna patterns.

Fig. 11 summarizes the extracted model parameters for all measured V2X antennas at 5915 MHz, including roof-, mirror-, glass-, and bumper-mounted configurations. Clear position-dependent trends can be observed. Roof-mounted antennas generally exhibit smaller values of $Q$, indicating weaker pattern distortion due to their relatively unobstructed radiation environment. In contrast, bumper- and glass-mounted antennas tend to exhibit larger $Q$ or lower $G_0$, reflecting stronger vehicle-body effects and less favorable installation conditions. These results demonstrate that the proposed parameters can effectively capture the impact of antenna mounting position on radiation behavior. Note that the proposed model is demonstrated and validated using V2X antennas as a representative example. Its extension to other frequency bands and wireless functions across the complete measurement dataset will be investigated in future work.

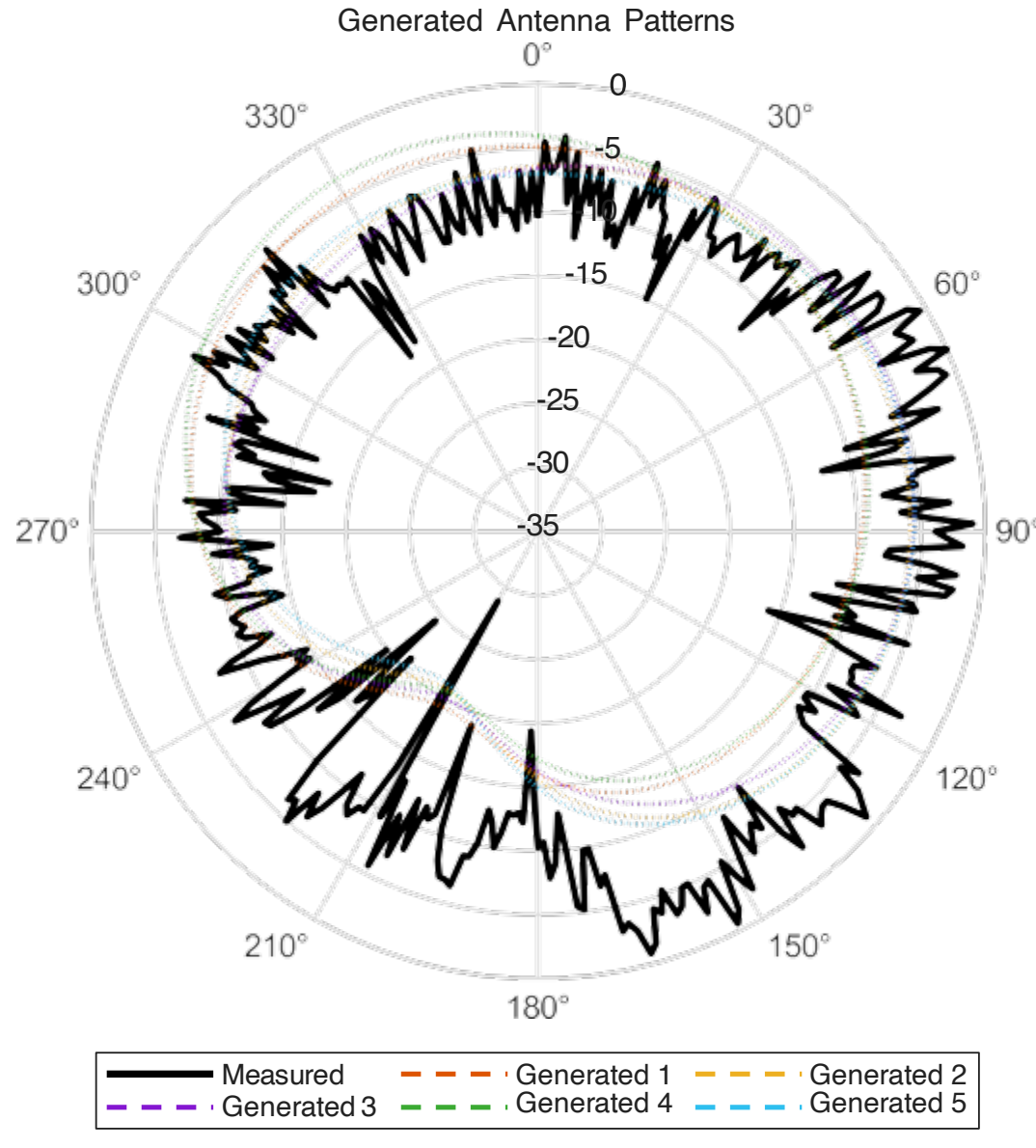

Fig. 10. Generated vehicle-mounted antenna patterns from realizations of the proposed model, for the bumper-mounted V2X antenna at 5915 MHz.

## V. CONCLUSIONS

This paper presented a large-scale experimental study of vehicle-mounted antenna radiation patterns based on full-vehicle SNF measurements. A comprehensive dataset covering ten commercial vehicle platforms, multiple antenna mounting locations, and various wireless functions (4G, 5G, V2X, and GNSS) covering the application frequencies, was established under a unified measurement framework. The measured results demonstrated that vehicle integration can significantly reshape antenna radiation patterns through blockage and structural asymmetry, leading to directional imbalance, localized radiation nulls, and function-dependent coverage characteristics. To quantitatively characterize these effects, a statistical framework based on HPR, PAG, UHE, and horizontal fluctuation metrics was applied to the measured dataset. The statistical analysis revealed distinct radiation characteristics among different wireless functions and installation configurations. Furthermore, a novel statistical model was developed to generate realistic vehicle-mounted antenna patterns using a small set of physically interpretable parameters. The generated patterns were shown to preserve the dominant characteristics of measured radiation patterns, providing a practical representation for system-level simulations. This work provides a large-scale empirical foundation for the characterization and modeling of realistic vehicle-mounted antenna radiation patterns, supporting realistic system-level simulation and system design for intelligent transportation applications.

## ACKNOWLEDGMENTS

This work was supported in part by the Tianjin Science and Technology Major Project (No. 25ZXCKQY00080), and in part by the HORIZON-MSCA NEDS-6G Project under Grant 101209281.

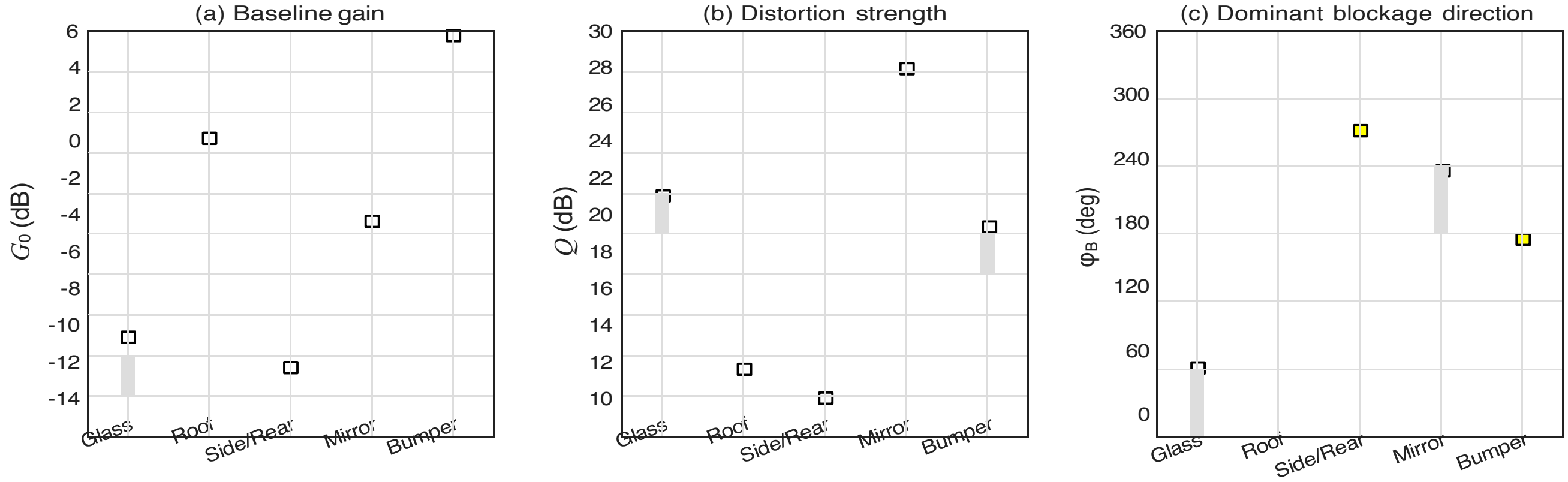

Fig. 11. Parameter values for model realization, extracted from the V2X antenna measurements at 5915 MHz.

## References

[1] M. Sierra-Castañer, "Recent developments in automotive antenna measurements," in *2020 14th European Conference on Antennas and Propagation (EuCAP)*, 2020, pp. 1–4.

[2] P. Pelland, D. J. van Rensburg, M. Berbeci *et al.*, "Automotive ota measurement techniques and challenges," in *2020 Antenna Measurement Techniques Association Symposium (AMTA)*, 2020, pp. 1–6.

[3] J. Wu, Y. Qi, P. Shen *et al.*, "Challenges and solutions for automotive ota testing," in *2023 IEEE Symposium on Electromagnetic Compatibility Signal/Power Integrity (EMC+SIPI)*, 2023, pp. 105–108.

[4] L. Durand, L. Duchesne, and L. J. Foged, "Fast antenna testing using advanced probe array technology," in *Proceedings of the Fourth European Conference on Antennas and Propagation*, 2010, pp. 1–4.

[5] Y. Jing, R. Chen, X. Yang *et al.*, "Analysis of vehicle body impact on antenna pattern and effective antenna aperture for full vehicle antenna testing," in *2025 19th European Conference on Antennas and Propagation (EuCAP)*, 2025, pp. 1–5.

[6] G. Artner, W. Kotterman, G. D. Galdo *et al.*, "Conformal automotive roof-top antenna cavity with increased coverage to vulnerable road users," *IEEE Antennas and Wireless Propagation Letters*, vol. 17, no. 12, pp. 2399–2403, 2018.

[7] T. Sayama, O. Kagaya, H. Shoji *et al.*, "Characteristics evaluation and path loss measurement of vehicle glass mounted antenna for 28-ghz band," in *2019 13th European Conference on Antennas and Propagation (EuCAP)*, 2019, pp. 1–5.

[8] M. E. Asghar, F. Wollenschläger, C. Bornkessel *et al.*, "Comparative analysis of spherical near-field automotive antenna measurement facilities," in *2019 13th European Conference on Antennas and Propagation (EuCAP)*, 2019, pp. 1–5.

[9] P. N. Betjes and D. Pototzki, "A hemispherical near-field system for automotive antenna testing," 2005. [Online]. Available: https://api.semanticscholar.org/CorpusID:173986213

[10] M. E. Asghar, F. Wollenschläger, A. Asgharzadeh *et al.*, "Influence of antenna mounting location on the radiation pattern of an automotive antenna," in *12th European Conference on Antennas and Propagation (EuCAP 2018)*, 2018, pp. 1–5.

[11] D. Kremer, A. Morris, R. Blake *et al.*, "Outdoor far-field antenna measurements system for testing of large vehicles," in *2012 6th European Conference on Antennas and Propagation (EUCAP)*, 2012, pp. 2256–2260.

[12] P. Noren, L. J. Foged, and P. Garreau, "State of the art spherical near-field antenna test systems for full vehicle testing," in *2012 6th European Conference on Antennas and Propagation (EUCAP)*, 2012, pp. 2244–2248.

[13] B. Derat, M. Celik, W. Simon *et al.*, "Optimization of in-vehicle connectivity through simulation-augmented antenna measurements," in *2022 Antenna Measurement Techniques Association Symposium (AMTA)*, 2022, pp. 1–6.

[14] R. Karli, M. Becquaert, and L. Bontemps, "Assessment of the antenna mounting location on a military vehicle for jamming applications based on computational electromagnetics," in *2023 International Conference on Military Communications and Information Systems (ICMCIS)*, 2023, pp. 1–8.

[15] M. G. Fernández, Y. López, and F. L.-H. Andrés, "Antenna measurement and diagnostics processing techniques using unmanned aerial vehicles," in *2019 13th European Conference on Antennas and Propagation (EuCAP)*, 2019, pp. 1–5.

[16] F. Hou, M. Chen, Y. Zhao *et al.*, "Antenna far-field measurement system based on software defined radio and uav," in *2025 IEEE International Workshop on Radio Frequency and Antenna Technologies (iWRFAT)*, 2025, pp. 1–5.

[17] P. O. Iversen, J. Estrada, F. Saccardi *et al.*, "Experimental comparison of vehicular antenna measurements performed over different floors," in *2020 14th European Conference on Antennas and Propagation (EuCAP)*, 2020, pp. 1–5.

[18] M. S. L. Mocker, M. Schiller, R. Brem *et al.*, "Combination of a full-wave method and ray tracing for radiation pattern simulations of antennas on vehicle roofs," in *2015 9th European Conference on Antennas and Propagation (EuCAP)*, 2015, pp. 1–5.

[19] S. Kim, D. Lee, B. Kim *et al.*, "On-vehicle integration and experimental validation of a glass-embedded antenna," in *Proceedings of the 2025 International Symposium on Antennas and Propagation (ISAP)*, 2025, session Number: Pos1; Publication Date: 2025-10-27; Online ISSN: 2188-5079.

[20] F. Saccardi, A. Scannavini, L. Scialacqua *et al.*, "Virtual drive testing based on automotive antenna measurements for evaluation of vehicle-to-x communication performances," in *2019 IEEE International Symposium on Antennas and Propagation and USNC-URSI Radio Science Meeting*, 2019, pp. 927–928.

[21] Y. Ji, W. Fan, M. G. Nilsson *et al.*, "Virtual drive testing over-the-air for vehicular communications," *IEEE Transactions on Vehicular Technology*, vol. 69, no. 2, pp. 1203–1213, 2020.

[22] F. Saccardi, F. Mioc, A. Scannavini *et al.*, "Experimental investigation of different floor materials in automotive near field antenna testing," in *2020 Antenna Measurement Techniques Association Symposium (AMTA)*, 2020, pp. 1–6.

[23] Z. Chen, "Sampling effect on radiation characteristics of automotive 5g antenna in spherical-near-field test range," in *2021 15th European Conference on Antennas and Propagation (EuCAP)*, 2021, pp. 1–5.

[24] N. Mutonkole, E. R. Samuel, D. I. L. de Villiers *et al.*, "Parametric modeling of radiation patterns and scattering parameters of antennas," *IEEE Transactions on Antennas and Propagation*, vol. 64, no. 3, pp. 1023–1031, 2016.

[25] J. Ø. Nielsen and G. F. Pedersen, "Mobile handset performance evaluation using radiation pattern measurements," *IEEE Transactions on Antennas and Propagation*, vol. 54, no. 7, pp. 2154–2165, 2006.

[26] A. Sibille, C. Roblin, S. Bories *et al.*, "A channel-based statistical approach to antenna performance in uwb communications," *IEEE Transactions on Antennas and Propagation*, vol. 54, no. 11, pp. 3207–3215, 2006.

[27] J. Ø. Nielsen, G. F. Pedersen, K. Olesen *et al.*, "Statistics of measured body loss for mobile phones," *IEEE Transactions on Antennas and Propagation*, vol. 49, no. 9, pp. 1351–1353, 2001.

[28] C. A. Balanis, *Antenna Theory: Analysis and Design*, 4th ed. Wiley, 2016.